\documentclass[aps,twocolumn,twoside,superscriptaddress]{revtex4}

\usepackage{dcolumn}
\usepackage{amsmath}

\usepackage{graphicx}

\def\O{{\rm O}}
\def\half{\mbox{$\frac12$}}

\begin{document}

\title{Community structure in social and biological networks}
\author{Michelle Girvan}
\affiliation{Santa Fe Institute, 1399 Hyde Park Road, Santa Fe, NM 87501}
\affiliation{Department of Physics, Cornell University, Clark Hall,
Ithaca, NY 14853--2501}
\author{M. E. J. Newman}
\affiliation{Santa Fe Institute, 1399 Hyde Park Road, Santa Fe, NM 87501}
\date{December 7, 2001}

\begin{abstract}
  A number of recent studies have focused on the statistical properties of
  networked systems such as social networks and the World-Wide Web.
  Researchers have concentrated particularly on a few properties which seem
  to be common to many networks: the small-world property, power-law degree
  distributions, and network transitivity.  In this paper, we highlight
  another property which is found in many networks, the property of
  community structure, in which network nodes are joined together in
  tightly-knit groups between which there are only looser connections.  We
  propose a new method for detecting such communities, built around the
  idea of using centrality indices to find community boundaries.  We test
  our method on computer generated and real-world graphs whose community
  structure is already known, and find that it detects this known structure
  with high sensitivity and reliability.  We also apply the method to two
  networks whose community structure is not well-known---a collaboration
  network and a food web---and find that it detects significant and
  informative community divisions in both cases.
\end{abstract}

\maketitle

\section{Introduction}
\label{intro}
Many systems take the form of networks, sets of nodes or vertices joined
together in pairs by links or edges~\cite{Strogatz01}.  Examples include
social networks~\cite{WF94,Scott00,WS98} such as acquaintance
networks~\cite{ASBS00} and collaboration networks~\cite{Newman01a},
technological networks such as the Internet~\cite{FFF99}, the World-Wide
Web~\cite{AJB99,Broder00}, and power grids~\cite{WS98,ASBS00}, and
biological networks such as neural networks~\cite{WS98}, food
webs~\cite{WM00}, and metabolic networks~\cite{Jeong00,FW00}.  Recent
research on networks among mathematicians and physicists has focused on a
number of distinctive statistical properties that most networks seem to
share.  One such property is the ``small world effect,'' which is the name
given to the finding that the average distance between vertices in a
network is short~\cite{PK78,Milgram67}, usually scaling logarithmically
with the total number $n$ of vertices.  Another is the right-skewed degree
distributions that many networks
possess~\cite{AJB99,Broder00,BA99,KRL00,DMS00}.  The degree of a vertex in a
network is the number of other vertices to which it is connected, and one
finds that there are typically many vertices in a network with low degree
and a small number with high degree, the precise distribution often
following a power-law or exponential form~\cite{Strogatz01,ASBS00,BA99}.

A third property that many networks have in common is clustering, or
network transitivity, which is the property that two vertices that are both
neighbors of the same third vertex have a heightened probability of also
being neighbors of one another.  In the language of social networks, two of
your friends will have a greater probability of knowing one another than
will two people chosen at random from the population, on account of their
common acquaintance with you.  This effect is quantified by the clustering
coefficient~$C$~\cite{WS98,NSW01}, defined by
\begin{equation}
C = {\mbox{$3\times$ (number of triangles on the graph)}\over
     \mbox{(number of connected triples of vertices)}}.
\label{defsc}
\end{equation}
This number is precisely the probability that two of one's friends are
friends themselves.  It is~1 on a fully connected graph (everyone knows
everyone else) and has typical values in the range $0.1$ to $0.5$ in many
real-world networks.

\begin{figure}[b]
\begin{center}
\resizebox{!}{5cm}{\includegraphics{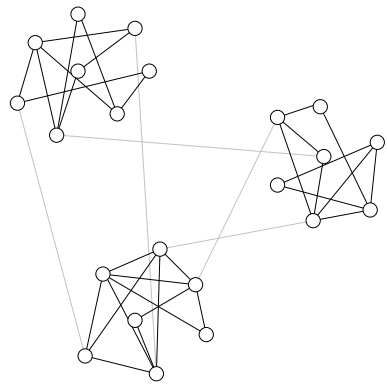}}
\end{center}
\caption{A schematic representation of a network with community structure.
  In this network there are three communities of densely connected vertices
  (circles with solid lines), with a much lower density of connections (gray
  lines) between them.}
\label{community}
\end{figure}

In this paper, we consider another property which, as we will show, appears
to be common to many networks, the property of community structure.  (This
property is also sometimes called clustering, but we refrain from this
usage to avoid confusion with the other meaning of the word clustering
introduced in the preceding paragraph.)  Consider for a moment the case of
social networks---networks of friendships or other acquaintances between
individuals.  It is matter of common experience that such networks seem to
have communities in them: subsets of vertices within which vertex--vertex
connections are dense, but between which connections are less dense.  A
figurative sketch of a network with such a community structure is shown in
Fig.~\ref{community}.  (Certainly it is possible that the communities
themselves also join together to form meta-communities, and that those
meta-communities are themselves joined together, and so on in a
hierarchical fashion.  This idea is discussed further in
Section~\ref{detecting}.)  The ability to detect community structure in a
network could clearly have practical applications.  Communities in a social
network might represent real social groupings, perhaps by interest or
background; communities in a citation network~\cite{Redner98} might
represent related papers on a single topic; communities in a metabolic
network might represent cycles and other functional groupings; communities
in the Web might represent pages on related topics.  Being able to identify
these communities could help us to understand and exploit these networks
more effectively.

In this paper we propose a new method for detecting community structure and
apply it to the study of a number of different social and biological
networks.  As we will show, when applied to networks for which the
community structure is already known from other studies, our method appears
to give excellent agreement with the expected results.  When applied to
networks for which we do not have other information about communities, it
gives promising results which may help us understand better the interplay
between network structure and function.

\section{Detecting community structure}
\label{detecting}
In this section we review existing methods for detecting community
structure and discuss the ways in which these approaches may fail, before
describing our own method, which avoids some of the shortcomings of the
traditional techniques.

\subsection{Traditional methods}
\begin{figure}[t]
\begin{center}
\resizebox{!}{3cm}{\includegraphics{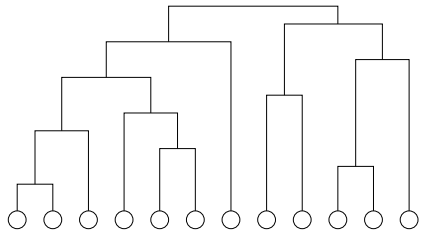}}
\end{center}
\caption{An example of a small hierarchical clustering tree.  The circles
  at the bottom of the figure represent the vertices in the network and the
  tree shows the order in which they join together to form communities
  for a given definition of the weight $W_{ij}$ of connections between
  vertex pairs.}
\label{hierarchical}
\end{figure}

The traditional method for detecting community structure in networks such
as that depicted in Fig.~\ref{community} is hierarchical clustering.  One
first calculates a weight $W_{ij}$ for every pair $i,j$ of vertices in the
network, which represents in some sense how closely connected the vertices
are.  (We give some examples of possible such weights below.)  Then one
takes the $n$ vertices in the network, with no edges between them, and adds
edges between pairs one by one in order of their weights, starting with the
pair with the strongest weight and progressing to the weakest.  As edges
are added, the resulting graph shows a nested set of increasingly large
components (connected subsets of vertices), which are taken to be the
communities.  Since the components are properly nested, they can all be
represented using a tree of the type shown in Fig.~\ref{hierarchical}, in
which the lowest level at which two vertices are connected represents the
strength of the edge which resulted in their first becoming members of the
same community.  A ``slice'' through this tree at any level gives the
communities which existed just before an edge of the corresponding weight
was added.  Trees of this type are sometimes called ``dendrograms'' in the
sociological literature.

Many different weights have been proposed for use with hierarchical
clustering algorithms.  One possible definition of the weight is the number
of node-independent paths between vertices.  Two paths which connect the
same pair of vertices are said to be node-independent if they share none of
the same vertices other than their initial and final vertices.  It is
known~\cite{Menger27} that the number of node-independent paths between
vertices $i$ and $j$ in a graph is equal to the minimum number of vertices
that need be removed from the graph in order to disconnect $i$ and $j$ from
one another.  Thus this number is in a sense a measure of the robustness of
the network to deletion of nodes~\cite{WH01}.

Another possible way to define weights between vertices is to count the
total number of paths that run between them (all paths, not just
node-independent ones).  However, since the number of paths between any two
vertices is infinite (unless it is zero), one typically weights paths of
length $\ell$ by a factor $\alpha^\ell$ with $\alpha$ small, so that the
weighted count of the number of paths converges~\cite{Katz53}.  Thus long
paths contribute exponentially less weight than short ones.  If ${\bf A}$
is the adjacency matrix of the network, such that $A_{ij}$ is~1 if there is
an edge between vertices $i$ and $j$ and~0 otherwise, then the weights in
this definition are given by the elements of the matrix
\begin{equation}
{\bf W} = \sum_{\ell=0}^\infty (\alpha{\bf A})^\ell
        = [{\bf I}-\alpha{\bf A}]^{-1}.
\end{equation}
In order for the sum to converge, we must choose $\alpha$ smaller than the
reciprocal of the largest eigenvalue of~${\bf A}$.

Both of these definitions of the weights give reasonable results for
community structure in some cases.  In other cases they are less
successful.  In particular, both have a tendency to separate single
peripheral vertices from the communities to which they should rightly
belong.  If a vertex is, for example, connected to the rest of a network by
only a single edge then, to the extent that it belongs to any community, it
should clearly be considered to belong to the community at the other end of
that edge.  Unfortunately, both the numbers of node-independent paths and
the weighted path counts for such vertices are small and hence single nodes
often remain isolated from the network when the communities are
constructed.  This and other pathologies, along with poor results from
these methods in some networks where the community structure is well known
from other studies, make the hierarchical clustering method, although
useful, far from perfect.

\subsection{Edge betweenness and community structure}
To sidestep the shortcomings of the hierarchical clustering method, we here
propose a new approach to the detection of communities.  Instead of trying
to construct a measure which tells us which edges are most central to
communities, we focus instead on those edges which are {\em least\/}
central, the edges which are most ``between'' communities.  Rather than
constructing communities by adding the strongest edges to an initially
empty vertex set, we construct them by progressively removing edges
from the original graph.

Vertex ``betweenness'' has been studied in the past as a measure of the
centrality and influence of nodes in networks.  First proposed by
Freeman~\cite{WF94,Freeman77}, the betweenness centrality of a vertex $i$
is defined as the number of shortest paths between pairs of other vertices
which run through~$i$.  It is a measure of the influence of a node over the
flow of information between other nodes, especially in cases where
information flow over a network primarily follows the shortest available
path.

In order to find which edges in a network are most ``between'' other pairs
of vertices, we generalize Freeman's betweenness centrality to edges and
define the {\bf edge betweenness} of an edge as the number of shortest
paths between pairs of vertices that run along it.  If there is more than
one shortest path between a pair of vertices, each path is given equal
weight such that the total weight of all the paths is unity.  If a network
contains communities or groups that are only loosely connected by a few
inter-group edges, then all shortest paths between different communities
must go along one of these few edges.  Thus, the edges connecting
communities will have high edge betweenness.  By removing these edges, we
separate groups from one another and so reveal the underlying community
structure of the graph.

The algorithm we propose for identifying communities is simply stated as
follows:
\begin{enumerate}
\setlength{\itemsep}{0pt}
\item Calculate the betweenness for all edges in the network.
\item Remove the edge with the highest betweenness.
\item Recalculate betweennesses for all edges affected by the removal.
\item Repeat from step 2 until no edges remain.
\end{enumerate}

As a practical matter, we calculate the betweennesses using the fast
algorithm of Newman~\cite{Newman01b}, which calculates betweenness for all
$m$ edges in a graph of $n$ vertices in time $\O(mn)$.  Since this
calculation has to be repeated once for the removal of each edge, the
entire algorithm runs in worst-case time~$\O(m^2n)$.  However, following
the removal of each edge, we only have to recalculate the betweennesses of
those edges that were affected by the removal, which is at most only those
in the same component as the removed edge.  This means that running time
may be better than worst-case for networks with strong community structure
(ones which rapidly break up into separate components after the first few
iterations of the algorithm).

To try to reduce the running time of the algorithm further, one might be
tempted to calculate the betweennesses of all edges only once and then
remove them in order of decreasing betweenness.  We find however that this
strategy does not work well, because if two communities are connected by
more than one edge, then there is no guarantee that all of those edges will
have high betweenness---we only know that at least one of them will.  By
recalculating betweennesses after the removal of each edge we ensure that
at least one of the remaining edges between two communities will always
have a high value.

\section{Tests of the method}
In this section we present a number of tests of our algorithm on
computer-generated graphs and on real-world networks for which the
community structure is already known.  In each case we find that our
algorithm reliably detects the known structure.

\subsection{Computer-generated graphs}
\begin{figure}
\begin{center}
\resizebox{!}{6cm}{\includegraphics{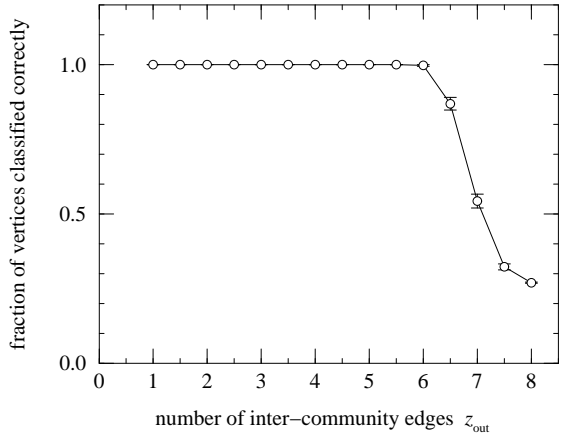}}
\end{center}
\caption{The fraction of vertices correctly classified by our method as the
number $z_{\rm out}$ of inter-community edges per vertex is varied, for
computer generated graphs of the type described in the text.  The
measurements with half-integer values $z_{\rm out}=k+\half$ are for graphs
in which half the vertices had $k$ inter-community connections and half
had~$k+1$.  Each point is an average over 100 realization of the graphs.
Lines between points are included solely as a guide to the eye.}
\label{rand_test}
\end{figure}

To test the performance of our algorithm on networks with varying degrees
of community structure, we have applied it to a large set of artificial,
computer-generated graphs similar to those depicted in
Fig.~\ref{community}.  Each graph was constructed with 128 vertices, each
of which was connected to exactly $z=16$ others.  The vertices were divided
into four separate communities with some number $z_{\rm in}$ of each
vertex's 16 connections made to randomly chosen members of its own
community and the remaining $z_{\rm out}=z-z_{\rm in}$ made to random
members of other communities.  This produces graphs which have known
community structure, but which are essentially random in other respects.
Using these graphs, we tested the performance of our algorithm as the ratio
of intra-community to inter-community connections was varied.  The results
are shown in Fig.~\ref{rand_test}.  As we can see, the algorithm performs
near perfectly when $z_{\rm out}\le6$, classifying virtually 100\% of
vertices into their correct communities.  Only for $z_{\rm out}>6$ does the
fraction correctly classified start to fall off.  In other words the
algorithm performs perfectly almost to the point at which each vertex has
as many inter-community connections as intra-community ones.  This is an
encouraging first result for the method.

\subsection{Zachary's karate club study}
\label{karate}
While computer-generated networks provide a reproducible and
well-controlled test-bed for our community-structure algorithm, it is
clearly desirable to test the algorithm on data from real-world networks as
well.  To this end, we have selected two datasets representing real-world
networks for which the community structure is already known from other
sources.  The first of these is drawn from the well-known ``karate club''
study of Zachary~\cite{Zachary77}.  In this study, Zachary observed 34
members of a karate club over a period of two years.  During the course of
the study, a disagreement developed between the administrator of the club
and the club's instructor, which ultimately resulted in the instructor's
leaving and starting a new club, taking about a half of the original club's
members with him.

\begin{figure}
\begin{center}
\resizebox{!}{10cm}{\includegraphics{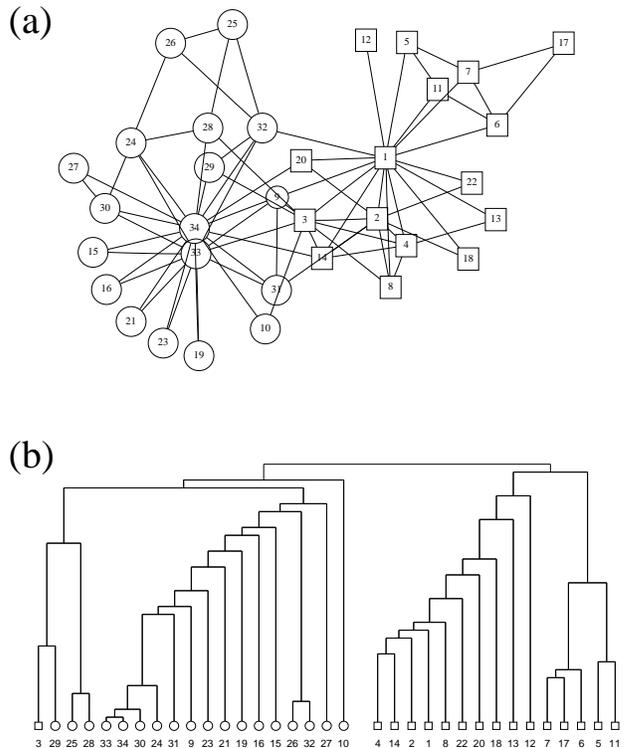}}
\end{center}
\caption{(a)~The friendship network from Zachary's karate club
  study~\cite{Zachary77}, as described in the text.  Nodes associated with
  the club administrator's faction are drawn as circles, while those
  associated with the instructor's faction are drawn as squares.  (b)~The
  hierarchical tree showing the complete community structure for the
  network.  The initial split of the network into two groups is in
  agreement with the actual factions observed by Zachary, with the
  exception that node~3 is misclassified.}
\label{zachary}
\end{figure}

Zachary constructed a network of friendships between members of the club,
using a variety of measures to estimate the strength of ties between
individuals.  Here we use a simple unweighted version of his network and
apply our algorithm to it in an attempt to identify the factions involved
in the split of club.  Figure~\ref{zachary}a shows the network, with the
instructor and the administrator represented by nodes 1 and 34,
respectively.  Figure~\ref{zachary}b shows the hierarchical tree of
communities produced by our method.  The most fundamental split in the
network is the first one at the top of the tree, which divides the network
into two groups of roughly equal size.  This split corresponds almost
perfectly with the actual division of the club members following the
break-up, as revealed by which club they attended afterwards.  Only one
node, node~3, is classified incorrectly.  In other words, the application
of our algorithm to the empirically observed network of friendships is a
good predictor of the subsequent social evolution of the group.

\subsection{College football}
\label{colleges}
As a further test of our algorithm, we turn to the world of US college
football.  (``Football'' here means American football, not soccer.)  The
network we look at is a representation of the schedule of Division~I games
for the 2000 season: vertices in the graph represent teams (identified by
their college names) and edges represent regular season games between the
two teams they connect.  What makes this network interesting is that it
incorporates a known community structure.  The teams are divided into
``conferences'' containing around 8 to 12 teams each.  Games are more
frequent between members of the same conference than between members of
different conferences, with teams playing an average of about 7
intra-conference games and 4 inter-conference games in the 2000 season.
Inter-conference play is not uniformly distributed; teams that are
geographically close to one another but belong to different conferences are
more likely to play one another than teams separated by large geographic
distances.

\begin{figure}
\begin{center}
\resizebox{!}{20cm}{\includegraphics{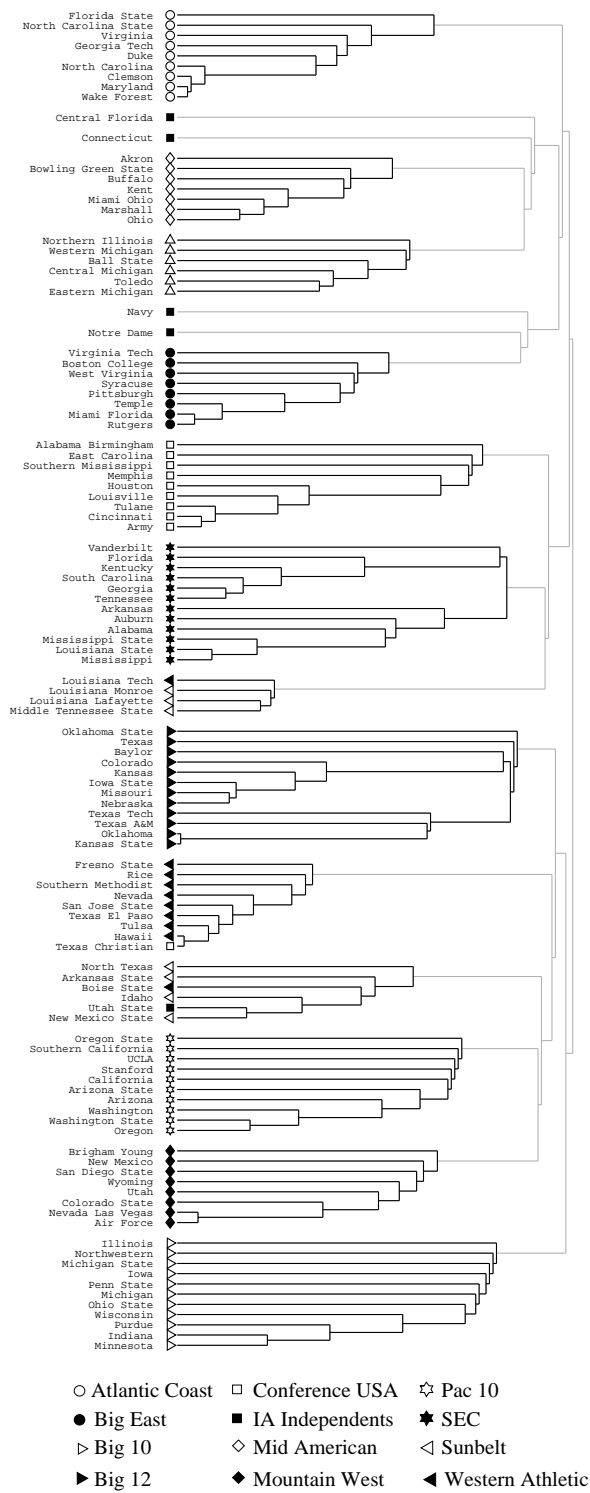}}
\end{center}
\caption{Hierarchical tree for the network reflecting the schedule of
regular season Division I college football games for year 2000.  Nodes in
the network represent teams and edges represent games between teams.  Our
algorithm identifies nearly all the conference structure in the network.}
\label{football}
\end{figure}

Applying our algorithm to this network, we find that it identifies the
conference structure with a high degree of success.  Almost all teams are
correctly grouped with the other teams in their conference.  There are a
few independent teams that do not belong to any conference---these tend to
be grouped with the conference with which they are most closely associated.
The few cases in which the algorithm seems to fail actually correspond to
nuances in the scheduling of games.  For example, the Sunbelt conference is
broken into two pieces and grouped with members of the Western Athletic
conference.  This happens because the Sunbelt teams played nearly as many
games against Western Athletic teams as they did against teams in their own
conference.
Naturally, our algorithm fails in cases like this where the network
structure genuinely does not correspond to the conference structure.  In
all other respects however it performs remarkably well.

\begin{figure}
\begin{center}
\resizebox{!}{9cm}{\includegraphics{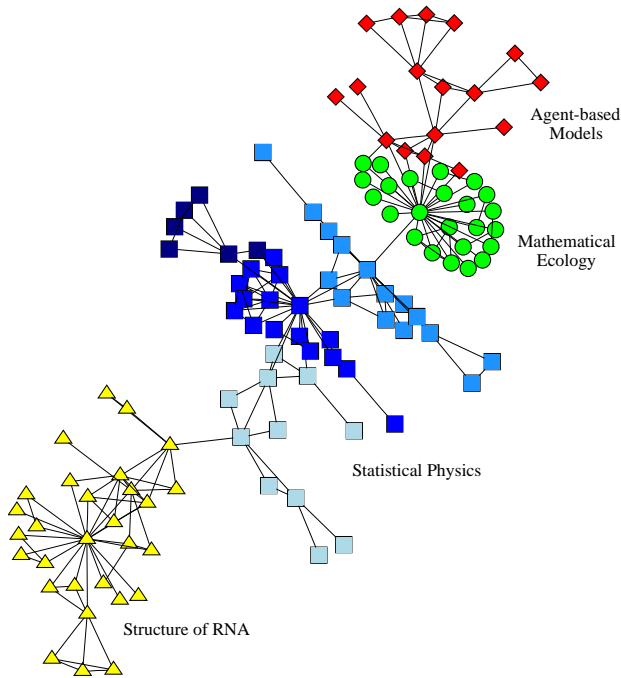}}
\end{center}
\caption{The largest component of the Santa Fe Institute collaboration
  network, with the primary divisions detected by our algorithm represented
  by different vertex shapes.}
\label{sfi}
\end{figure}

\section{Applications} 
\label{applications}
In the previous section we tested our algorithm on a number of networks for
which the community structure was known beforehand.  The results indicate
that our algorithm is a sensitive and accurate method for extracting
community structure from both real and artificial networks.  In this
section, we apply our method to two more networks for which the structure
is not known, and show that in these cases it can help us to understand the
make-up of otherwise complex and tangled datasets.  Our first example is a
collaboration network of scientists; our second is a food web of marine
organisms in the Chesapeake Bay.

\subsection{Collaboration network}
We have applied our community-finding method to a collaboration network of
scientists at the Santa Fe Institute, an interdisciplinary research center
in Santa Fe, New Mexico (and current academic home to both the authors of
this paper).  The 271 vertices in this network represent scientists in
residence at the Santa Fe Institute during any part of calendar year 1999
or 2000, and their collaborators.  An edge is drawn between a pair of
scientists if they coauthored one or more articles during the same time
period.  The network includes all journal and book publications by the
scientists involved, along with all papers that appeared in the institute's
technical reports series.  On average, each scientist coauthored articles
with approximately five others.

In Fig.~\ref{sfi} we illustrate the results from the application of our
algorithm to the largest component of the collaboration graph (which
consists of 118 scientists).  Vertices are drawn as different shapes
according to the primary divisions detected.  We find that the algorithm
splits the network into a few strong communities, with the divisions
running principally along disciplinary lines.  The community at the top of
the figure (diamonds) is the least well defined, and represents a group of
scientists using agent-based models to study problems in economics and
traffic flow.  The algorithm further divides this group into smaller
components that correspond roughly with the split between economics and
traffic.  The next community (circles) represents a group of scientists
working on mathematical models in ecology, and forms a fairly cohesive
structure, as evidenced by the fact that the algorithm does not break it
into smaller components to any significant extent.  The largest community
(represented by the squares) is a group working primarily in statistical
physics, and is sub-divided into several well-defined smaller groups which
are denoted by the various shadings.  In this case, each sub-community
seems to revolve around the research interests of one dominant member.  The
final community at the bottom of the figure (triangles) is a group working
primarily on the structure of RNA.  It too can be divided further into
smaller sub-communities, centered once again around the interests of
leading members.

Our algorithm thus seems to find two types of communities: scientists
grouped together by similarity either of research topic or of methodology.
It is not surprising to see communities built around research topics; we
expect scientists to collaborate primarily with others with whom their
research focus is closely aligned.  The formation of communities around
methodologies is more interesting, and may be the mark of truly
interdisciplinary work.  For example, the grouping of those working on
economics with those working on traffic models may seem surprising, until
one realizes that the technical approaches these scientists have taken are
quite similar.  As a result of these kinds of similarities, the network
contains ties between researchers from traditionally disparate fields.  We
conjecture that this feature may be peculiar to interdisciplinary centers
like the Santa Fe Institute.

\subsection{Food web}
We have also applied our algorithm to a food web of marine organisms living
in the Chesapeake Bay, a large estuary on the east coast of the United
States.  This network was originally compiled by Baird and
Ulanowicz~\cite{Baird89} and contains 33 vertices representing the
ecosystem's most prominent taxa.  Most taxa are represented at the species
or genus level, although some vertices represent groups of related species.
Edges between taxa indicate trophic relationships---one taxon feeding on
another.  Although relationships of this kind are inherently directed, we
here ignore direction and consider the network to be undirected.

\begin{figure}
\begin{center}
\resizebox{!}{7cm}{\includegraphics{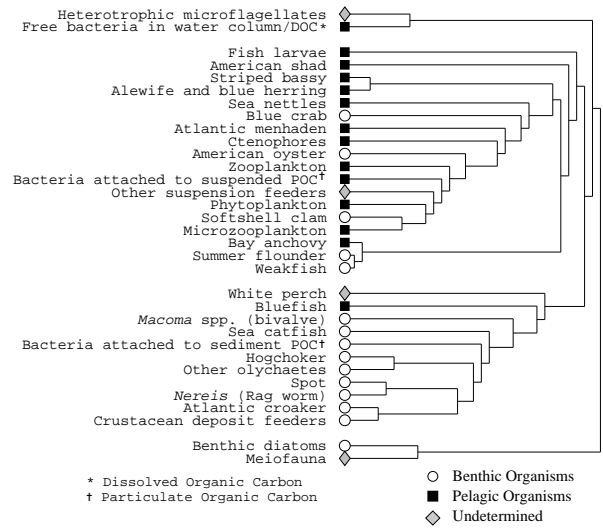}}
\end{center}
\caption{Hierarchical tree for the Chesapeake Bay food web described in the
text.}
\label{foodweb}
\end{figure}

Applying our algorithm to this network, we find two well-defined
communities of roughly equal size, plus a small number of vertices that
belong to neither community---see Fig.~\ref{foodweb}.  As the figure shows,
the split between the two large communities corresponds quite closely with
the division between pelagic organisms (ones that dwell principally near
the surface or in the middle depths of the bay) and benthic organisms (ones
that dwell near the bottom).  Interestingly, the algorithm includes within
each group organisms from a variety of different trophic levels.  This
contrasts with other techniques that have been used to analyze food
webs~\footnote{J. J. Luczkovich, personal communication.}, which tend to
cluster taxa according to trophic level rather than habitat.  Our results
seem to imply that pelagic and benthic organisms in the Chesapeake Bay can
be separated into reasonably self-contained ecological sub-systems.  The
separation is not perfect: a small number of benthic organisms find their
way into the pelagic community, presumably indicating that these species
play a substantial role in the food chains of their surface-dwelling
colleagues.  This suggests that the simple traditional division of taxa
into pelagic or benthic may not be an ideal classification in this case.

We have also applied our method to a number of other food webs.
Interestingly, while some of these show clear community structure similar
to that of Fig.~\ref{foodweb}, some others do not.  This could be because
some ecosystems are genuinely not composed of separate communities, but it
could also be because many food webs, unlike other networks, are dense,
i.e.,~the number of edges scales as the square of the number of vertices
rather than scaling linearly~\cite{Martinez92}.  Our algorithm was designed
with sparse networks in mind, and it is possible that it may not perform as
well on dense networks.

\section{Conclusions} 
\label{concs}
In this paper we have investigated community structure in networks of
various kinds, introducing a new method for detecting such structure.
Unlike previous methods which focus on finding the strongly connected cores
of communities, our approach works by using information about edge
betweenness to detect community peripheries.  We have tested our method on
computer generated graphs and have shown that it detects the known
community structure with a high degree of success.  We have also tested it
on two real-world networks with well-documented structure and find the
results to be in excellent agreement with expectations.  In addition, we
have given two examples of applications of the algorithm to networks whose
structure was not previously well-documented and find that in both cases it
extracts clear communities which appear to correspond to plausible and
informative divisions of the network nodes.

A number of extensions or improvements of our method may be possible.
First, we hope to generalize the method to handle both weighted and
directed graphs.  Second, we hope that it may be possible to improve the
speed of the algorithm.  At present, the algorithm runs in time $\O(n^3)$
on sparse graphs, where $n$ is the number of vertices in the network.  This
makes it impractical for very large graphs.  Detecting communities in, for
instance, the large collaboration networks~\cite{Newman01a} or subsets of
the Web graph~\cite{Broder00} that have been studied recently, would be
entirely unfeasible.  Perhaps, however, the basic principles of our
approach---focusing on the boundaries of communities rather than their
cores, and making use of edge betweenness---can be incorporated into a
modified method that scales more favorably with network size.

We hope that the ideas and methods presented here will prove useful in the
analysis of many other types of networks.  Possible further applications
range from the determination of functional clusters within neural networks
to analysis of communities on the World-Wide Web, as well as others not yet
thought of.  We hope to see such applications in the future.

\begin{acknowledgments}
  The authors would like to thank Jennifer Dunne, Neo Martinez, Matthew
  Salganik, Steve Strogatz, and Doug White for useful conversations, and
  Jennifer Dunne, Sarah Knutson, Matthew Salganik, and Doug White for help
  compiling the data for the food web, collaboration, college football, and
  karate club networks, respectively.  This work was funded in part by the
  National Science Foundation under grant number DMS--0109086.
\end{acknowledgments}

\end{document}